\documentstyle[12pt]{article}
\date {}
\def\be{\begin{equation}}
\def\ee{\end{equation}}
\begin{document}
\begin{center}
{\bf Angular Velocity Operator and Barnett-Pegg Formalism }\\
{ Ramandeep S. Johal}\\ {\it Department of
Physics,}\\ {\it Panjab University, Chandigarh - 160014,
India.}\\ email: raman\%phys@puniv.chd.nic.in
\end{center}
\begin{abstract}
We define a new operator within Barnett-Pegg formalism of phase
angle. The physical predictions for this operator correspond to
those expected of angular velocity operator. Examples studied
are particle on a ring with and without magnetic field and
quantum harmonic oscillator.
\end{abstract}
\newpage
Difficulties in the proper understanding of phase operator in
quantum mechanics [1] have proportinately hindered the
comprehension of angular velocity operator also, which is, in
general, defined as time rate of change of the phase operator.
Recently, Pegg-Barnett formalism [2] was reasonably successful
in elucidating the phase operator for quantum harmonic
oscillator [3,4]. Barnett and Pegg also extended it to the study
of rotation angle operator [5]. However, when applied to
calculate the expectation values of angular velocity over
angular momentum states, it yields unphysical results [6]. Thus
the very significance of the angular velocity operator being
talked in this formalism was put to question.
More recently,
Abe suggested that Barnett-Pegg (BP) formalism may be looked as
an inherently $q$-deformed theory [7]. To remove the criticism
of Loss and Mullen [6], a $q$-deformed Heisenberg equation of
motion for rotation angle operator was used. Then in the limit
of large-$l$, in which the angle operator becomes continuous,
one obtains a physically acceptable result.

In this letter, we
give a new definition of angular velocity operator within BP
formalism, which gives physical results in the large-$l$ limit.
We also apply the same definition to phase of quantum harmonic
oscillator as well as charged particle on a ring in the presence
of magnetic field.

The physical system underlying BP formalism
is a particle of mass M moving on a circle of radius R. The
rotation angle state is defined in $(2l+1)$-dimensinal Hilbert
$H^{2l+1}$, as follows:
\be
|\theta_n\rangle=\frac{1}{\sqrt{2l+1}}\sum^{l}_{m=-l}
e^{-im\theta_n}|m\rangle
\qquad n=0,1,...,2l. \label{thet}
\ee
where $\theta_n = \theta_0 + \frac{2\pi n}{2l+1}$ and $m$ is
eigenvalue of the $z$-component of angular momentum :
$\hat{L}_z|m\rangle=m|m\rangle, (m=-l,-l+1,...,l).$ We take
$\theta_0=0$ for simplicity.  It would not affect the results on
angular velocity. The deformation parameter may be introduced as
$q=e^{{-i}\frac{2\pi}{2l+1}}$.

Now $\{|\theta_n\rangle\}$ and
$\{|m\rangle\}$ are dual representations. Note the discrete
displacement property of operators on them:
\be
e^{i\hat {\Phi}}|m\rangle=|m+1\rangle \qquad (m\ne l)
\ee
\be
e^{i\hat {\Phi}}|l\rangle=|-l\rangle
\ee
whose dual relations are
\be
q^{\hat{L}_z}|\theta_n\rangle=|\theta_{n+1}\rangle
\ee
\be
q^{\hat{L}_z}|\theta_{2l}\rangle=|\theta_0=0\rangle
\ee
We can also have $e^{-i\hat {\Phi}}$ and $q^{-\hat{L}_z}$
operators, which displace the respective states in direction
opposite to above. We regard $e^{i\hat {\Phi}}$ and
$q^{\hat{L}_z}$ as the natural pair of dual operators in BP
formalism. In fact these operators $q$-commute
\be
[q^{\hat{L}_z},e^{i\hat {\Phi}}]_q= q^{\hat{L}_z}e^{i\hat
{\Phi}} -qe^{i\hat {\Phi}}q^{\hat{L}_z}=0
\ee
Now as remarked in [6], $\frac{d\hat {\Phi}}{dt}$ does not lead
to physical results for angular velocity.  In the following, we
seek a new operator, defined in terms of the unitary phase
operator and show that its expectation values correspond well
with those expected from a genuine angular velocity operator.
Define
\be
\hat{R} = -ie^{-i\hat
{\Phi}}\frac{d}{dt}e^{i\hat {\Phi}} \label{angv}
\ee
Although standard Heisenberg equation does not work for $\hat
{\Phi}$ (within BP formalism), nothing restricts us to use the
same for unitary phase operator. Thus we write
\be
\frac{d}{dt}e^{i\hat {\Phi}}= \frac{1}{i\hbar}[e^{i\hat
{\Phi}},H]  \label{Hes}
\ee
Strictly speaking, the time derivative in the definition
(\ref{angv}) should be a discrete time dervative as our state
space is discrete and finite dimensional.  But as we do not use
deformed Heisenberg equation ( as was done in [7]) so the final
relation is not affected.  Using (\ref{Hes}) in (\ref{angv}) we
get
\be
\hat{R}=\frac{1}{\hbar}(e^{-i\hat
{\Phi}}He^{i\hat {\Phi}}- H).
\ee
Clearly, $\hat{R}$ is hermitian as well as
periodic.

The Hamiltonian of the rotating particle is given by
\be
H_0=\frac{{\hat L}^{2}_{z}{\hbar}^2}{2MR^2}.
\ee
Therefore
\be
\langle l|\hat{R}|l\rangle = \frac{(2l+1)\hbar}{2MR^2}.
\ee
In the semiclassical limit of large-$l$, the r.h.s of the above
relation is equal to $\frac{\hbar l}{MR^2}$, which is in accord
with Ehrenfest theorem.

Next, we calculate the rate of change of the phase operator for
quantum harmonic oscillator as given in [2]. The phase state in
$(s+1)$-dimensional Hilbert space is given by
\be
|\theta_m\rangle=\frac{1}{\sqrt{s+1}}\sum^{s}_{n=0}
q^{mn}|m\rangle
\qquad m=0,1,...,s.
\ee
where $q=e^{-i\frac{2\pi}{s+1}}$.

The unitary phase operator acts as shift operator on the number
states:
\be
e^{i\hat {\Phi}}|n\rangle = |n+1\rangle.
\ee
The relevant hamiltonian is given by $H=(\hat N + \frac{1}{2}
)\hbar \omega$.  Applying eq. (9) to the harmonic oscillator, we
get
\be
\langle n|\hat{R}|n\rangle = \omega .
\ee
This result is same for all number states. This ensures that in
the limit of $s\to \infty$, we get the expected physical result.

Finally, we consider a charged particle of mass $M$, charge $e$,
on a circle of radius $R$, subjected to a weak magnetic field
through the center of the ring. Suppose, ${\bf B} = B\hat z$.
The hamiltonian for the system is
\be
H = H_0 - \frac{e}{2Mc}B{\hat L}_z.
\ee
$H_0$ is given by eq. (10). Then the expectation values of
$\hat{R}$ calculated over eigenstates of ${\hat
{L}}_z$ show that angular velocity is decreased by an amount
$\frac{eB}{2Mc}$, which is equal to the Larmour frequency. The
change will be of opposite sign for particle rotating in
opposite direction. (Note that these results are meaningful only
in the infinite limit, in which only the phase angle becomes
continuous.

Concluding, we have defined a new operator which serves well as
an analogue of angular velocity in Barnett-Pegg formalism. This
is corroborated with some simple examples.
\vskip 12pt
\section*{References}
\begin{enumerate}
\item {} P. Carruthers and M.M. Nieto, Rev. Mod. Phys. {\bf 40}
(1968) 411 and references therein.
\item {} D.T. Pegg and S.M. Barnett, Europhys. Lett. {\bf 6}
(1988) 483; Phys. Rev. A {\bf 39} (1989) 1665.
\item {} J.A. Vaccaro, S.M. Barnett and D.T. Pegg, J. Mod. Opt.
{\bf 39} (1992) 603.
\item {} C. Wagner, R.J. Brecha, A. Schenzle and H. Walther,
Phys. Rev. A {\bf 46} (1992) R5350.
\item {} S.M. Barnett and D.T. Pegg, Phys. Rev. A {\bf 41}
(1990) 3427.
\item {} D. Loss and K. Mullen, J. Phys. A:Math and Gen. {\bf
25} (1992) L235.
\item {} S. Abe, Phys. Rev. A {\bf 54} (1996) 93.
%\item {} S. Abe, Phys. Lett. A {\bf 200} (1995) 239.
\end{enumerate}
\end{document}